\documentclass{article}

\usepackage{arxiv}
\usepackage[utf8]{inputenc} 
\usepackage[T1]{fontenc}    
\usepackage{url}            
\usepackage{booktabs}       
\usepackage{amsfonts}       
\usepackage{microtype}      

\usepackage{graphicx}
\usepackage{amssymb}
\usepackage{float}
\usepackage{eurosym}

\usepackage{mathtools}
\usepackage{ccicons}
\usepackage[numbers]{natbib}
\usepackage[
    pdfauthor={Pietro Zambelli – pzambelli@eurac.edu},
    pdftitle={Integrated approach for the identification of spatial patterns related to renewable energy potential in European territories},
    pdfsubject={This study was conducted in the frame of the European Union (EU) H2020 project HotMaps (grant number: 723677) project. 
The Horizon 2020 HotMaps project (https:\/\/www.hotmaps-project.eu) aims at developing "an open source heating/cooling mapping and planning toolbox" for EU28 policy makers at national and local level. 
This article is published in: Scaramuzzino et al. 2019 Integrated approach for the identification of spatial patterns related to renewable energy potential in European territories. Renewable and Sustainable Energy Reviews, 101:1 – 13, 2019. DOI: https:\/\/doi.org\/10.1016\/j.rser.2018.10.024.
This work is licensed under a Creative Commons Attribution-NonCommercial-NoDerivatives 4.0 International License.
The study presents an effort to classify the territories of a specific area, according to similarities in the estimated potential of their renewable sources, considering also their economic and sociodemographic structure and their geographic features.
Specifically, the paper focuses on the area of EU28 and Switzerland and uses as basis for the analysis, data estimating the potential of renewable energy sources collected and elaborated in the framework of the project HotMaps (Horizon 2020).
The method used to group the territorial units is cluster analysis, and specifically the k-means algorithm. The data present some interesting patterns and the territories of EU28 and Switzerland at NUTS3 level are classified into 17 clusters. 
The analysis shows the presence of heterogeneity within national borders and among territories comprised in the macro regions target of specific EU programmes, specifically the Adriatic-Ionian region, the Alpine region, the Baltic Sea region and the Danube region.
The results of this research are meant to be used by European policy makers in developing more focused transnational renewable energy policies and strategies.},
    pdfkeywords={Renewable energy sources; Energy planning; Energy policy; EU28},
            ]{hyperref}

\makeatletter
\providecommand{\doi}[1]{
  \begingroup
    \let\bibinfo\@secondoftwo
    \urlstyle{rm}
    \href{http://dx.doi.org/#1}{
      doi:\discretionary{}{}{}
      \nolinkurl{#1}
    }
  \endgroup
}
\makeatother

\newcounter{savecntr} 
\newcounter{restorecntr} 

\title{Integrated approach for the identification of spatial patterns related to renewable energy potential in European territories}

\author{
  Chiara Scaramuzzino\setcounter{savecntr}{\value{footnote}}\thanks{
          \href{http://www.eurac.edu}{Eurac Research} ---
          \href{http://www.eurac.edu/en/research/technologies/renewableenergy}{Institute of Renewable Energy}: 
          \href{http://www.eurac.edu/en/research/technologies/renewableenergy/researchfields/Pages/Energy-strategies-and-planning.aspx}{Urban and Regional Energy System Group}
      }
  \\
  \texttt{\href{mailto:chiara.scaramuzzino@eurac.edu}{cscaramuzzino@eurac.edu}} \\
  \And
  Giulia Garegnani\setcounter{restorecntr}{\value{footnote}}%
      \setcounter{footnote}{\value{savecntr}}\footnotemark
      \setcounter{footnote}{\value{restorecntr}}
      \thanks{Corresponding author}
  \\
  \texttt{\href{mailto:giulia.garegnani@eurac.edu}{ggaregnani@eurac.edu}} \\
  \And
  Pietro Zambelli\setcounter{restorecntr}{\value{footnote}}%
      \setcounter{footnote}{\value{savecntr}}\footnotemark
      \setcounter{footnote}{\value{restorecntr}}
  \\
  \texttt{\href{mailto:pietro.zambelli@eurac.edu}{pzambelli@eurac.edu}} \\
}

\date{Accepted by:\\ 
\textbf{Renewable and Sustainable Energy Reviews}\thanks{
    \citet{scaramuzzino_integrated_2019} \href{http://www.sciencedirect.com/science/article/pii/S1364032118307275}{Integrated approach for the identification of spatial patterns related to renewable energy potential in European territories}. \textit{Renewable and Sustainable Energy Reviews}, 101:1 – 13, 2019. DOI: \href{https://doi.org/10.1016/j.rser.2018.10.024}{10.1016/j.rser.2018.10.024}.
\newline
    This work is licensed under a \href{https://creativecommons.org/licenses/by-nc-nd/4.0/}{Creative Commons Attribution-NonCommercial-NoDerivatives 4.0 International License} \ccbyncnd.
   }
\\
20$^{th}$ Oct 2018}

\begin{document}
\maketitle

\begin{abstract}
The study presents an effort to classify the territories of a specific area, according to similarities in the estimated potential of their renewable sources, considering also their economic and sociodemographic structure and their geographic features.
Specifically, the paper focuses on the area of EU28 and Switzerland and uses as basis for the analysis, data estimating the potential of renewable energy sources collected and elaborated in the framework of the project HotMaps (Horizon 2020).
The method used to group the territorial units is cluster analysis, and specifically the k-means algorithm. The data present some interesting patterns and the territories of EU28 and Switzerland at NUTS3 level are classified into 17 clusters. 
The analysis shows the presence of heterogeneity within national borders and among territories comprised in the macro regions target of specific EU programmes, specifically the Adriatic-Ionian region, the Alpine region, the Baltic Sea region and the Danube region.
The results of this research are meant to be used by European policy makers in developing more focused transnational renewable energy policies and strategies.
\end{abstract}


\section{Introduction}
\label{intro}

Energy production and use represent a large share of greenhouse gas emissions responsible for global warming and climate change.
Replacing fossil fuels with renewable sources of energy provides the opportunity to tackle these phenomena by limiting the increase of global temperature.
In 2014, the European Council renewed its commitment to the energy transition through the 2030 Energy Strategy, which, among other goals, aims at reaching a target of 27\% for the share of renewables in energy consumption within 2030 \cite{scarlat_renewable_2015}.\\

EU leaders have been putting in place a structured framework of policies and guidelines encouraging renewable energy production in order to reach the ambitious goals of the 2030 Energy Strategy \cite{directive2009_28,directive2016}. 
These policies do not have the same outcomes in all European countries, and this could happen for two reasons.
First, each country of EU28 has different policies, planning cultures and instruments for the promotion of renewable energies \cite{reiche_policy_2004}; when investigating the conditions for an effective promotion of renewable energies, \citet{reiche_policy_2004} found some Member States to be more successful than others in promoting renewable energies. 
On another level, a reason for some territories to be more successful in promoting renewable energy than other could reside in the fact that EU energy programmes are often put in place either at national level or within the boundaries of macro-regions, although both countries and macro-regions often comprise very diverse territories responding to policies in different ways. 
In order to support consistent outcomes from renewable energy promotion policies across Europe, two elements are crucial: promoting an effective exchange of information about successful conditions and practices and inform the implementation of EU energy policy with a more comprehensive knowledge of the territories at local level.

At the same time, the increase in renewable energy production challenges scientists and researchers to deal with the assessment of natural resource availability in European territories.
In the literature, several works \cite{bodis_could_2014,suri_potential_2007,magagna_ocean_2015,ericsson_assessment_2006, de_wit_european_2010,nadai_wind_nodate,estima_global_2013} investigated renewable energy potential in the EU.
Some studies \cite{suri_potential_2007,estima_global_2013} deal with raster data about energy potential at high spatial resolution, while others \cite{bodis_could_2014,magagna_ocean_2015,ericsson_assessment_2006, de_wit_european_2010} assess the potential of renewable energy sources at country level.
These approaches compute the energy potential starting from physical and geographic data.
The methodologies differ according to the assumptions for the definition of the energy potential. 
According to \citet{GAREGNANI2018709}, energy potential figures can refer to:
\begin{itemize}
    \item Theoretical potential when no constraints are taking into account but only physical laws are considered
    \item Planning potential if spatial, legal, social and environmental constraints are applied 
    \item Technical potential if the technology (i.e. conversion factors, energy losses, etc.) is considered
    \item Financial potential when the total energy potential of a region accounts for the financial assessment of different renewable energy plants.
\end{itemize}
The concept of energy potential has to do mainly with a theoretical and technical concept, without involving financial and planning features, when the spatial extent is wide, e.g. \cite{bodis_could_2014, suri_potential_2007, yohanis_geographic_2006}. 
\citet{de_vries_renewable_2007} provide a study of the global energy potential converting the technical potential to an economic output by estimating the production cost.
Instead, local studies account also for a sustainable use of energy sources and$/$or  technical and financial constraints by considering local parameters \cite{sacchelli_biomasfor:_2013, palomino_cuya_gis-based_2013, GAREGNANI2018709, zambelli_gis_2012,vettorato_spatial_2011}. 
The goal of energy potential assessment strictly depends on the applied definitions. Local studies, e.g \cite{sacchelli_biomasfor:_2013, palomino_cuya_gis-based_2013, GAREGNANI2018709}, provide geographical detailed information to administrators for energy planning. Technical, financial and planning constraints are considered. 
Instead, coarser estimations of energy potential \cite{bodis_could_2014, suri_potential_2007, de_vries_renewable_2007} define indicators supporting energy strategies at national scale.
Understanding and harmonizing the definition of energy potential becomes crucial when different renewable energy sources are compared.  \\

Although various assessment of renewable energy potential have been performed at European level, to our knowledge, no studies in the literature focus on the identification of spatial patterns enhancing the effectiveness of energy policies, yet.
Disciplines such as medicine and biology \cite{saxena_review_2017} use cluster analysis for spatial pattern identification, while fewer studies are available in planning and policy research.
The most relevant examples in the latter fields are reported as follows. 
The projects realized in the framework of ESPON, a program aiming at inspiring policy making with territorial evidence, tried to increase the effectiveness of energy policies by identifying and understanding various territorial aspects of specific administrative areas. 
For instance, the 2005 ESPON "Regional Classification of Europe" project aimed at providing policy makers with "information on territorial structures in Europe" through clusterization of municipalities based on socio-economic, geo-ecological and agricultural data \cite{pecher_typology_2013}.
\citet{metzger_climatic_2005} used PCA (Principal Component Analysis) and cluster analysis based on geomorphological variables to obtain a \textit{"statistical stratification of the European environment"}.
Cluster analysis based on economic variables has been employed by \citet{monfort_real_2013} to test the presence of real convergence, \textit{"the process whereby the GDP per capita levels of lower-income
economies catch up with those of higher-income economies on a durable basis"} \cite{ecb_real_2015}, in the EU.
\citet{pecher_typology_2013} tried to structure a "regional typology" of the Alpine Space, by using hierarchical cluster analysis based on 26 "spatial-pattern indicators" to classify the municipalities of the area. 
Though being thorough, this study considers exclusively geographic, topographic and landscape-related indicators thus its result is a classification of the territories in these terms.
Finally \citet{zhang_quantitative_2018} use attribute construction, PCA and k-means clustering to highlight temporal and spatial patterns related to wind and solar energy sources in the Shandong province in China with the aim of supporting renewable energy planning. 

The use of cluster analysis for the identification of energy related spatial patterns, especially in an area as vast as Europe, remains open to further investigation and an integrated approach which couples a comprehensive assessment of renewable energies with socioeconomic and geographic indicators is missing. 
Such an approach can lead to the finding of spatial patterns in the distribution of energy potential and help identify territorial heterogeneities within European countries and macro-regions as well as cross-borders similarities which can foster territorial cohesion. 

The evidence of energy related spatial patterns can support policy makers in defining plans and strategies for renewable energy production.

Since the aim of this study is to provide a tool for policy makers, the analysis can not be limited to an assessment of renewable energy potential: effective energy strategies call for a comprehensive knowledge of the territory where they need to be carried out.
This work proposes a classification of territories of EU28 and Switzerland based on energy potential, socioeconomic and geomorphological indicators by means of the \textit{k-means} cluster algorithm.
Energy potential indicators are presented in section \ref{sec:pot_selection} and \ref{sec:energy_indicators}.
Geographic and environmental features of the territories as well as indicators describing their economic and sociodemographic structure are presented in section \ref{sec:non_energy_indicators}. 

The cluster methodology is presented in section \ref{sec:cluster}.
Finally, the classification resulting from the analysis is presented and discussed in section \ref{results} with the help of GIS (Geographic Information System) tools, which allow the spatial visualization of results.\\

\section{Materials and methods}
\label{method}

\begin{figure}[!ht]
\centering\includegraphics[width=1\linewidth]{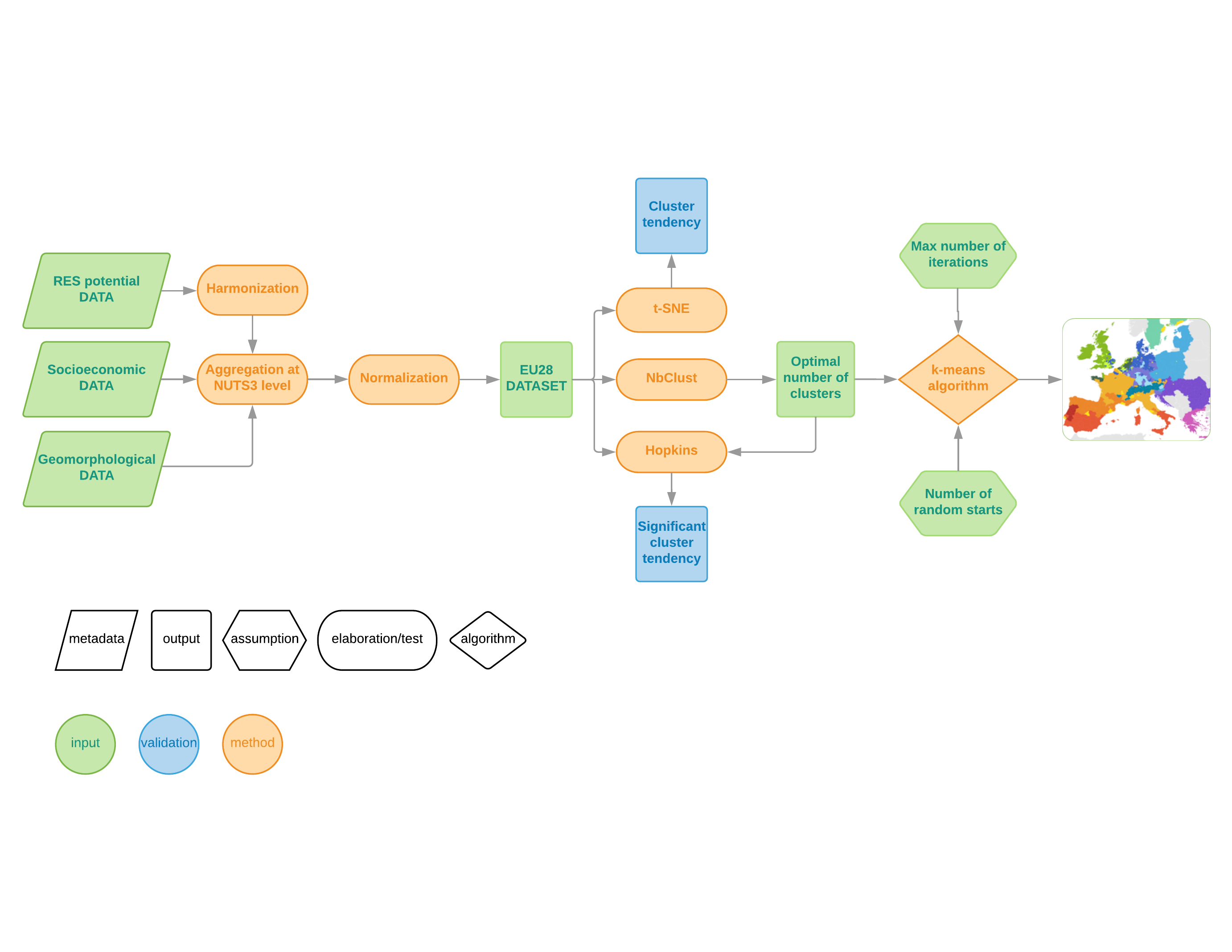}
\caption{Flowchart: the diagram shows the method applied by the authors starting from the harmonization and aggregation at NUTS3 level of all indicators, following with the normalization of data and tests over the presence of clusters and ending with the application of the cluster algorithm to the dataset.}
\label{flowchart}
\end{figure}

Two challenges arise when considering renewable energy potential indicators from different literature sources: the presence of different concepts of energy potential and a heterogeneity of spatial resolutions at which this potential is calculated.
For what concerns the use of an unambiguous energy potential concept, this study works with planning potentials, according to the definition reported in the previous section.
As for the second issue, until now, not much effort has been made towards leveling out the differences in spatial resolutions and analyze the data for multiple energy sources at the same geographic/administrative level. 
In this sense, this study fills a significant gap by harmonizing the spatial resolutions at which the potential of selected renewable sources has been calculated, in order to use this harmonized potential as a starting point for cluster analysis.
The methodology is described in the following sections and visualized through a flowchart in Figure \ref{flowchart}.

\subsection{Identification of an adequate territorial unit within the study area}
\label{sec:nuts}

In order to perform a classification of territories, the first step is finding an adequate territorial unit. 
Said territorial unit needs to fulfill two basic requirements:
\begin{itemize}
\item reliable data for the unit should be available or at least easily computable
\item the unit should be well known in the field of application and usable by the recipients of the results of the analysis: policy makers, advisors and other stakeholders from the public and private sector must be able to grasp the concept and visualize the administrative boundaries of the territorial unit in question and easily retrieve its energy policy framework.
\end{itemize}

The Classification of Territorial Units for Statistics (NUTS) is a \textit{"hierarchical system for dividing up the economic territory of the EU"} \cite{eurostat_nuts_2015}. 
According to this system, countries are divided based on their administrative partitions and the classification is structured in four levels
\begin{itemize}
\item NUTS0 is the less aggregated level, representing the country
\item NUTS1 usually represents states or macro-regions
\item NUTS2 typically represents regions
\item NUTS3 is the most aggregated level of administrative subdivision and depending on the country can indicate provinces (Italy, Spain), counties (Hungary, Croatia, Lithuania), districts or metropolitan areas (Germany, UK) and other types of administrative micro-regions.
\end{itemize}

The territorial unit chosen for this study is NUTS3, the most aggregated unit thus potentially the one showing most accurate and locally usable results. 
This approach presents some issues. 
First, the division at NUTS3 level draws the administrative subdivision of the territories differently depending on the country. 
Due to this fact, NUTS3 territories can be very diverse: for instance, in terms of surface area, the smallest NUTS3 unit, the city of Melilla in Spain, comprises an area of 13 km\textsuperscript{2}, while the biggest unit, the county of Norrbotten in Sweden, comprises an area of $10.5 \times 10^4$ km\textsuperscript{2}.
Secondly, energy policies are not always implemented at NUTS3 level, but rather at less aggregated levels.
Nevertheless, the NUTS3 division remains the most reliable when it comes to data availability with high spatial resolutions.
Furthermore, EU policy makers have been pushing forward with the collection of data at NUTS3 level, especially for energy related projects (ESPON ECGT and HOTMAPS \cite{noauthor_hotmaps_nodate}) thus data at NUTS3 level can be found in official databases or, when present at less aggregated levels, they can be easily re-elaborated.
As for the second requirement, the administrative borders of NUTS3 territorial units should be well-known in urban and regional energy planning: policy makers are expected to be familiar with this classification and if not, to quickly grasp its concept and context. 

Being the aim of the study the investigation of possible patterns in the distribution of renewable energy potential across a territory, the choice of indicators for this analysis is based on their relevance in the study area and on the availability of data at NUTS3 level.\\

\subsection{Selection of renewable energy sources and review of their potential}
\label{sec:pot_selection}

A second challenge lies with the interpretation of different definitions of renewable energy potential in existing datasets as shown in the introduction.
Using energy potential data as variables in a cluster analysis avoids the ranking of the different energy sources and partially solves this problem.
However, the introduction of specific assumption in the energy potential definition (e.g. protected areas, distances from urban areas, etc...) could lead to different spatial distribution affecting the results of the cluster analysis. 
Consequently, an effort has to be made in order to have similar concepts of potential for all energy sources in each territorial unit.

The data were collected from four main sources: the Eurostat database \cite{european_commission_database_nodate}, the Copernicus Land Monitoring Service \cite{eu_copernicus_nodate}, several European projects (cited in table \ref{tab_data} according to the renewable source) and elaborations carried out by the authors within the HotMaps Horizon2020 project. 

\citet{resch_potentials_2008} defined the global energy potential for biomass, geothermal, hydro-power, ocean, solar and wind energy.
Among the renewable sources proposed by \citet{resch_potentials_2008} biomass, solar and wind energy were selected for this study, while hydro-power and geothermal energy were not included in the analysis.  

In fact, \citet{paish_small_2002} highlighted as most of the hydro-power potential in Europe has been already exploited by dams and reservoir plants.
Besides, there are potentially conflicts between hydro-power production and the European Water Frame Directive \cite{directive2000directive} willing to preserve the water quality. 

Secondly, the potential from geothermal and aerothermal heat pump systems was excluded, due to the fact that this potential is strictly correlated with building design and technology and these systems need to use external power coming from the national energy mix.
Though recognizing the role these sources could play in covering the EU heating and cooling demand \cite{eucomunication}, they were excluded given the focus of this study being to assess a theoretical potential more than a technical one.

Regarding ocean wave potential, \citet{magagna_ocean_2015} analyzed status and future perspective in Europe.
However, due to the difficulty with finding an indicator for this energy potential at NUTS3 level, since strictly connected to the ocean area measured only at national level, this source was excluded from the study.

Data on the potential of agricultural and forest biomass and livestock effluents were collected in the framework of the Biomass Policies project by \citet{elbersen_outlook_2014}. 
The project analyzed data from several biomass sources also considering those cultivated purposely for biofuel production (energy crops). 
Energy crops were excluded from this study due to the debated environmental impacts in terms of land use change \cite{wise_agriculture_2014}, water resources depletion and biodiversity losses \cite{pacetti_waterenergy_2015, manzone_energy_2016}.
In this work, agricultural residues considered suitable for energy production were selected among those included in the report after subtracting from the available residues \cite{elbersen_outlook_2014}:
\begin{itemize}
\item the biomass to be left on the field for soil conservation
\item the biomass destined to conventional competing uses (feedstock production, livestock bedding, other agricultural and agro-industrial processes)
\end{itemize} 
Agricultural residues considered in the study are summarized in table \ref{tab_Agricultural residues}. 
The livestock effluents considered for the estimation of energy potential consist in solid and liquid manure from cattle, pigs and poultry. 

\begin{table}[h!]
\begin{footnotesize}
\centering
\begin{tabular}{l l l}
    \hline
    \textbf{Crop}                  & \textbf{Production process} & \textbf{Residue} \\
    \hline
    Cereals (excl. maize and rice) & Cereal production           & Straw            \\
    Maize                          & Maize production            & Stover           \\
    Oilseed rape and sunflower     & Oil production              & Stubble          \\
    Sugar beet                     & Sugar production            & Leaves and tops  \\
    Rice                           & Rice production             & Straw            \\
    Olives                         & Oil production              & Pits             \\
    Olives                         & Olive and oil production    & Pruning          \\
    Citrus                         & Citrus production           & Pruning          \\
    Grape                          & Wine production             & Pruning          \\
    \hline
\end{tabular}
\caption{Agricultural residues considered in the estimation of energy potential in relation to specific crops and production processes.}
\label{tab_Agricultural residues}
\end{footnotesize}
\end{table}

For what concerns the potential of waste water \citet{hepbasli_key_2014} highlights how this source can be exploited for energy extraction.
The potential of waste water is preferred over geothermal and aerothermal being less dependent on building design and possible to assess at district level.
Data on the energy potential from waste water was retrieved and elaborated in the framework of the Horizon2020 Hotmaps project \cite{noauthor_hotmaps_nodate}.

Municipal solid waste were included in the assessment, due to the role waste currently plays in energy extraction \cite{noauthor_srren_nodate}. 
In this study, for sustainability reasons, only the non recyclable share of urban waste is considered suitable for energy production.
2014 data on municipal solid waste from households and economic activities (Statistical Classification of Economic Activities in the European Community \cite{eurostat_statistical_nodate}) were collected from the Eurostat Database. 

Wind and solar power have been deeply investigated by several authors and within numerous projects, due to their acknowledged relevance in renewable energy production \cite{noauthor_srren_nodate}.
For what concerns wind power, only on-shore wind energy potential is considered here since there are no tools mapping the spatial feasibility of off-shore plants at European level.
Data on on-shore wind energy potential were collected from the Global Wind Atlas \cite{Global_Wind_Atlas,estima_global_2013}, a mapping tool managed by Irena and the DTU containing data on wind power density for 50, 100 and 200 m high hubs in the form of a raster layer with a  resolution. 
In order to estimate the theoretical potential of solar power, data on the solar radiation on optimally inclined surfaces were retrieved from the PVGIS project \cite{jrcs_institute_for_energy_and_transport_about_2012,huld_new_2012} in the form of a raster layer with a 1km x 1km resolution.

Table \ref{tab_data} refers to the data availability for the selected energy sources. 
All data refer to theoretical potentials: technical aspects (e.g. technology used, efficiency, energy losses, etc.) are not included in this assessment. 
Planning constraints were considered for each energy source, based on the assumptions made in the projects/studies assessing their potential.

\begin{table}[h!]
\begin{footnotesize}
\centering
\begin{tabular}{l l l l}
    \hline
    \textbf{Energy source}               & \textbf{Data source}                                            & \textbf{Spatial resolution} & \textbf{Proxy}               \\
    \hline
    Agricultural residues $P_{agr}$      & Intelligent Energy Europe \cite{elbersen_outlook_2014}          & NUTS0                       & LUCAS \cite{eu_lucas_nodate} \\
    Forest residues $P_{for}$            & Intelligent Energy Europe \cite{elbersen_outlook_2014}          & NUTS0                       & Corine Land Cover            \\
    Livestock residues $P_{liv}$         & Intelligent Energy Europe \cite{elbersen_outlook_2014}          & NUTS0                       & Eurostat                     \\
    Municipal solid waste $P_{mun}$      & Intelligent Energy Europe \cite{elbersen_outlook_2014}          & NUTS0                       & Eurostat                     \\
    Waste water treatment plant    $P_{ww}$ & Hotmaps Project \cite{noauthor_hotmaps_nodate}                  & Plant position              & -                            \\
    Wind $P_{wind}$                      & DTU Global Atlas \cite{Global_Wind_Atlas}                       & 1km x 1km                   & -                            \\
    Solar power    $P_{sun}$                & PVGIS \cite{jrcs_institute_for_energy_and_transport_about_2012} & 1km x 1km                   & -                            \\
    \hline
\end{tabular}
\caption{Data sources, original spatial resolution of data and proxy used to scale the energy potential at NUTS3 level for all energy sources.}
\label{tab_data}
\end{footnotesize}
\end{table}

\subsection{Harmonization of selected energy potential indicators at NUTS3 level}
\label{sec:energy_indicators}

In order to identify spatial patterns considering an homogeneous administrative unit, existing data on energy potential, available at different spatial resolutions, see table \ref{tab_data}, needed to be aggregated at NUTS3 level. 
Furthermore, all datasets needed to be harmonized to avoid considering different energy potential definitions in the cluster analysis.
For instance, while assessing the potential of biomass \cite{elbersen_outlook_2014} already introduced some planning and sustainability constraints, the data on wind and solar energy potential only regard the theoretical potential. 
Given the impossibility to consider legislative and local constraints at this spatial extent, this work uses quasi-planning potentials by considering some land use related criteria.
Consequently, in order to harmonize these wind and solar datasets (1km x 1km raster layers) with the biomass ones, these potential were re-elaborated by integrating GIS data into the original energy-potential dataset. 
Besides, as already mentioned in \ref{sec:nuts}, NUTS3 areas can comprise very different territories, in terms of areas, population and land covers.
For this reason, in order to obtain an indicator of energy potential considering the differences among territories, the potentials of all renewable sources were normalized, as described for each source and summarized in table \ref{tab_energy_potential}.
Furthermore it is more interesting for the analysis carried out here to have an idea of the resource mix per unit (area or population) more than the total potential from RES per region. \\

Biomass data, including agricultural and forest residues and municipal solid waste, were available at national level, thus required a down-scaling at NUTS3 level by means of GIS tools.
Data on the energy potential of agricultural biomass were aggregated at NUTS3 level by using the LUCAS framework \cite{eu_lucas_nodate}, a survey that provides statistics on land use and cover in the EU28 territory. 
The LUCAS grid was re-elaborated to extract only the land cover classes relevant to our research, i.e. those relative to the selected biomasses, for each NUTS3 unit.
Data on the energy potential of livestock effluents were redistributed at NUTS3 level using the Eurostat statistics on the number of manure storage facilities as proxy. 
Forest biomass here accounts only for two categories of residues originated from forest management, namely, fuel wood, and residues from fuel wood and round wood.
The energy potential of these residues was redistributed using as proxy the Corine Land Cover classification. 
The Corine Land Cover is a 100m x 100m raster layer that classifies land covers into 44 categories. 
The use of Corine Land Cover could be a simplification in some countries where the nature of forest cover can change a lot depending on the latitude (Italy, Spain, France), but an estimation of the forest potential is enough in order to perform the cluster. To our knowledge there are no more accurate dataset at EU level available to use as proxy for distributing the forest biomass potential at NUTS3 level.
The categories related to forest cover (broad-leaved forest, coniferous forest, mixed forest, natural grassland, moors and heathland, sclerophyllous vegetation, transitional woodland shrub) were used to redistribute the potential on the territory of each country. 
This potential was then aggregated at NUTS3 level.
All biomass potentials were normalized by dividing them by the area of the territory.\\

The potential of municipal solid waste from household and economic activities $\hat{P}_{mun}$ at national level was computed by multiplying the quantity of available waste $Q_{mun}$ by a lower heating value ($l_h$) of 13,8 MJ Kg$^{-1}$ and an equivalence ratio (r) of 0,3 \cite{arena_process_2012}, according to the following formula:
\begin{align*}
\label{msw}
  \hat{P}_{mun} = Q_{mun} \cdot l_h \cdot  r \\
\end{align*}
The 2011 Census from the Eurostat Database \cite{european_commission_database_nodate} was used as proxy to redistribute the potential of household waste at NUTS3 level. 
Assuming that the production of waste is proportional to the production of goods and services (represented by the GDP) in a region, The 2014 GDP statistics from the Eurostat Database \cite{european_commission_database_nodate} was used as proxy to redistribute the potential of waste from economic activities at NUTS3 level. 
The potential from waste was normalized by divided it by the population of each NUTS3 area. \\

Water treatment plant potential was calculated in the framework of the Horizon2020 HotMaps project \cite{noauthor_hotmaps_nodate}.
Waste water treatment plants were divided in suitable or not suitable for energy production based on their capacity and proximity to urban areas \cite{neugebauer_mapping_2015}: these planning criteria were followed in order to obtain a realistic measure of the availability of the resource.
Once identified the suitable plants, their power was aggregated at NUTS3 level.
Similarly to municipal solid waste, the potential from waste water was normalized by dividing it by the population of each NUTS3 area.\\

The wind planning potential was estimated by integrating these planning criteria \cite{hastik_using_2016} in GIS:
\begin{itemize}
\item Using only areas with low or sparse vegetation, bare and burnt areas (extracted from the Corine Land Cover)
\item Excluding 1 km buffer areas containing cities and settlements (extracted from the Corine Land Cover)
\item Excluding areas above 2500 m.a.s.l.
\item Excluding corridors for bird connectivity (Common Database on Designated Areas, CDDA)
\item Excluding protected areas of the Natura2000 network
\item Considering a distance among wind hubs of 300 m
\end{itemize}
The wind potential at NUTS3 level is represented by the annual median value per wind hub in each territory. \\

In order to estimate a feasible solar energy potential, the calculations were made assuming to install solar panels only on building roofs: the European Settlement Map, a raster layer downloaded from the Copernicus Land Monitoring System \cite{eu_copernicus_nodate} with a 10m x 10m resolution, served as building footprint.
The solar potential at NUTS3 level is represented by the annual median value per square meter in each territory.\\ 

\begin{table}[h!]
\begin{footnotesize}
\centering
\begin{tabular}{l l l}
    \hline
    \textbf{Indicator} & \textbf{Measure}                                             & \textbf{Unit of measure} \\
    \hline
    $\hat{P}_{agr}$    & Potential of agricultural residues per km\textsuperscript{2} & PJ km$^{-2}$             \\
    $\hat{P}_{for}$    & Potential of forest residues per km\textsuperscript{2}       & PJ km$^{-2}$             \\
    $\hat{P}_{liv}$    & Potential of livestock residues per km\textsuperscript{2}    & PJ km$^{-2}$             \\
    $\hat{P}_{mun}$    & Potential of municipal solid waste per person                & PJ inhab$^{-1}$          \\
    $\hat{P}_{ww}$     & Potential of waste water per person                          & KW inhab$^{-1}$          \\
    $\hat{P}_{wind}$   & Potential of wind at 100 meters per wind hub                 & GWh hub$^{-1}$           \\
    $\hat{P}_{sun}$    & Potential of solar power per m\textsuperscript{2}            & kWh m$^{-2}$             \\
    \hline
\end{tabular}
\caption{Energy potential indicators, normalized measure and unit for the selected sources.}
\label{tab_energy_potential}
\end{footnotesize}
\end{table}

\subsection{Choice and harmonization of non-energy territorial indicators at NUTS3 level}
\label{sec:non_energy_indicators}
The exploitation of the potential from renewable sources strictly depends on specific territorial aspects.
\citet{del_rio_assessing_2008} highlight how renewable energy production effects local development, above all in rural area, by creating industries and employment opportunities. 
\citet{belmonte_potential_2009} stress the importance of the connections between energy potential and socio-environmental conditions for the creation of energy related territorial plans. 
For this reasons, it seemed essential to include some sort of indicators about the spatial location, geographic features and economic and sociodemographic structure of the territorial units.

Universally acknowledged indicators, like surface area, population and Gross Domestic Product were first considered.
Electricity and gas prices and data on heating and cooling degree days were included to give a rough estimation of the energy demand of territories: \citet{quayle_heating_1979} showed a very high correlation between energy consumption and heating degree days. 
Starting from the monthly average of heating and cooling degree days, annual data were obtained.
As shown by \citet{yohanis_real-life_2008} energy demand depends also on specific variables describing the type of dwellings and occupancy.  
This work does not deal with the estimation of energy consumption for a specific country. 
For this reason, a proxy, such as heating and cooling degree days, was used as indicator of the energy demand. 
The resulting indicators provided information about the need of a territory for heating and cooling and allowed to perform a more accurate classification.

Most indicators were collected from the Eurostat database and redistributed at NUTS3 level when available only at lower spatial resolutions (see Table \ref{tab:indicator} for details). 
Indicators available at NUTS2 level were redistributed simply by using the same value reported for lower spatial resolution since the concerned indicators were either expressed in percentage, in function of the population or as daily values.
Electricity and gas prices, the only two indicators available at NUTS0, were treated the same way, since expressed in euros per kWh.

The indicator representing the percentage of urban areas at NUTS3 level was obtained by extracting from the Corine Land Cover the classes related to urban areas and calculating the percentage of those areas on the total surface of each territory. 

The indicator measuring elevation is based on EU-DEM, a digital surface model with a spatial resolution of 25m x 25m made available by Copernicus Land Monitoring Service \cite{eu_copernicus_nodate-1}, and the measure taken was the median value at NUTS3 level.

Finally, in order to include a measure for spatial closeness among NUTS3 areas, the coordinates of the centroid of each territory were used to create the indicators latitude and longitude.
Notice that the spatial closeness of centroid coordinates does not guarantee the spatial contiguity and the connection between NUTS3 units. 
Even if connectivity is a relevant indicator concerning networks and flows, spatial closeness partially describes similarities among territories and potential cooperation.

\begin{table}[h!]
\begin{footnotesize}
\centering
\begin{tabular}{l l l}
    \hline
    \textbf{Indicator} & \textbf{Measure}                                                       & \textbf{Unit of measure} \\
    \hline
    $lat$              & Latitude of centroid                                                   & deg                      \\
    $long$             & Longitude of centroid                                                  & deg                      \\
    $A$                & Surface area                                                           & km$^{2}$                 \\
    $\rho_{pop}$       & Population density                                                     & inhab km$^{-2}$          \\
    $N_{pop}$          & Population                                                             & inhab                    \\
    $A_{urban}$        & Urban areas                                                            & \%                       \\
    $gdp$              & Gross Domestic Product                                                 & \EUR                     \\
    $p_{el}$           & Electricity price $^{*}$                                               & \EUR\  kWh $^{-1}$       \\
    $p_{gas}$          & Gas price $^{*}$                                                       & \EUR\  kWh $^{-1}$       \\
    $R_{edu}$          & Percentage of inhabitants with tertiary education $^{**}$              & \%                       \\
    $R_{unemp}$        & Percentage of unemployed people between 20 and 64    $^{**}$              & \%                       \\
    $income$           & Disposable income of private households per capita    $^{**}$             & \EUR\ inhab$^{-1}$       \\
    $gdp_{pps}$        & Gross Domestic Product in Purchasing Power Standard per person    $^{**}$ & \EUR\ inhab$^{-1}$       \\
    $r\&d$             & Research and Development expenditure per capita    $^{**}$                & \EUR\ inhab$^{-1}$       \\
    $h$                & Surface elevation                                                      & m a.s.l.                 \\
    $hdd$              & Annual heating degree days     $^{**}$                                    & K d$^{-1}$               \\
    $cdd$              & Annual cooling degree days $^{**}$                                     & K d$^{-1}$               \\
    \hline
\end{tabular}
\caption{Non-energy related indicators, measure, with indication of initial spatial resolution when different than NUTS3 ($^{*}$ at NUTS0 level and $^{**}$ at NUTS2) and unit.}
\label{tab:indicator}
\end{footnotesize}
\end{table}

\subsection{A cluster analysis to classify territories based on renewable energy potential}
\label{sec:cluster}
Cluster analysis encompasses different mathematical methods to group objects according to similarities: objects characterized by similar features are gathered in the same cluster according to a given algorithm \cite{romesburg_cluster_2004}.
Besides merely classifying a number of objects, cluster analysis offers interesting cues for testing research hypotheses \cite{romesburg_cluster_2004}. 
There are different variations of cluster analysis: the one used in this study is the \textit{k-means}, a non-hierarchical clustering method. 
This method is chosen over others due to its simplicity: \textit{k-means} is not only fast and intuitive to use, but it is also easy to explain to the recipients of the results. 
The primary intent of this study is to provide public authorities a tool for renewable energy planning. 
In order to make the results understandable and usable, the analysis performed to obtain them must be clear enough to people with no statistical or mathematical background, as energy managers and planners, local administrators and policy makers. \\

The \textit{k-means} algorithm selects \textit{k} objects to use as cluster seeds, and then distributes the remaining objects to the clusters.
The distribution into clusters is based on the Euclidean distance between every two objects: the following formula defines this distance.
\begin{align*}
  d(x,y) &= \left(\sum_{j=1}^{d} (x_j - y_i)^2\right)^\frac{1}{2} \\
\end{align*}
The cluster seeds are chosen in order to have a large Euclidean distance between them while the distribution of the remaining objects is done by keeping the Euclidean distance within the same cluster as small as possible \cite{romesburg_cluster_2004}.

The first step in cluster analysis is the selection of the variables \cite{anderberg_cluster_2014}. 
This step is particularly delicate since, as \citet{kaufman_introduction_1990} stated, in cluster analysis \textit{"a variable not containing any relevant information is worse than useless, because it will make the clustering less apparent"}. 
The occurrence of so-called \textit{trash variables} has the potential to cripple the whole clustering process because \textit{"they yield a lot of random terms in distances, thereby hiding the useful information provided by the other variables"} \cite{anderberg_cluster_2014}.
All variables contribute equally to define the cluster structure.

The cluster analysis was carried out in R \cite{R}, using the energy and non-energy indicators as variables. 

The database containing observations related to all EU28 NUTS3 areas was re-organized and the presence of missing data was checked. 
Variables presenting more than 18\% of missing values were excluded from the analysis. 
In variables presenting missing values below 18\% of the total, these were replaced by the mean value for that variable. 
A second test checked the presence of correlation between the variables: a correlation greater than the absolute value of 0.85 was tested.

In order to perform cluster analysis, and specifically the \textit{k-means} algorithm, data needed to be standardized. 
This step was necessary since each variable ranges over a different scale, with different orders of magnitude. 

Once the data were on the same scale for all variables, the presence of natural clusters in the dataset, i.e. the tendency of the data to group, was tested. 

First, the cluster tendency in the distribution of the data was tested by linear multidimensionality reduction through the t-distributed stochastic neighbor embedding (t-SNE) algorithm. 
t-SNE is not an exact method for determining the optimal number of clusters present in a dataset, it rather gives a rough estimation of it. 

The second tool used to estimate the number of clusters possibly present in the data was \textit{NbClust} \cite{charrad_nbclust:_2014}. 
This algorithm uses 30 indices for determining the best number of clusters, by also giving the user the possibility to choose three key arguments
\begin{itemize}
\item    the measure of the distance used to compute the dissimilarity matrix
\item    the minimum and the maximum number of clusters 
\item    the cluster analysis method
\end{itemize}
In this case, the distance measure employed was Euclidean, the minimum and maximum number of clusters were set respectively at 15 and 20 and the method chosen was \textit{k-means}.  
With these arguments, the number of clusters proposed by the highest number of indices was 17.

The last test used to confirm the tendency of the data to group in 17 clusters was the Hopkins statistic \cite{banerjee_validating_2004}. 
The result of the Hopkins statistic is a number from zero to 1: the closer the number to zero, the higher the tendency of the data to cluster in the given numbers of groups; with values far below 0.5, the dataset is significantly clusterable. 
The Hopkins statistic with 17 clusters gave a result of $6.95 \cdot 10^{-2}$ , confirming a significant tendency of the data to group into the optimal number of clusters given by the \textit{NbClust} algorithm.

After confirming the cluster tendency and having possibly found the best number of clusters for the dataset, the \textit{k-means} algorithm was applied to the data. 
This algorithm clusters the data matrix, in this case the scaled dataset, according to a given number of clusters (17), a maximum number of iterations (50), and a number of random starts (multiple random starts are recommended, we used 10), and finally returns:
\begin{itemize}
\item    the number of elements comprised in each cluster
\item    the centroid of each cluster for all variables
\item    a vector pairing each observation with the cluster it belongs to
\item     the variance between the clusters over the total variance in the data
\end{itemize}
The last information is particularly relevant: it indicates how much the variance in the dataset is explained by the clustering. 
The higher this percentage, the more significant the clustering of the data based on the selected variables: in this case, this figure was 61,3\%. 
It can be assumed that a significant percentage of the dataset variance is explained by the clustering.

\section{Results and discussion}
\label{results} 

\begin{figure}[!ht]
\centering\includegraphics[width=1\linewidth]{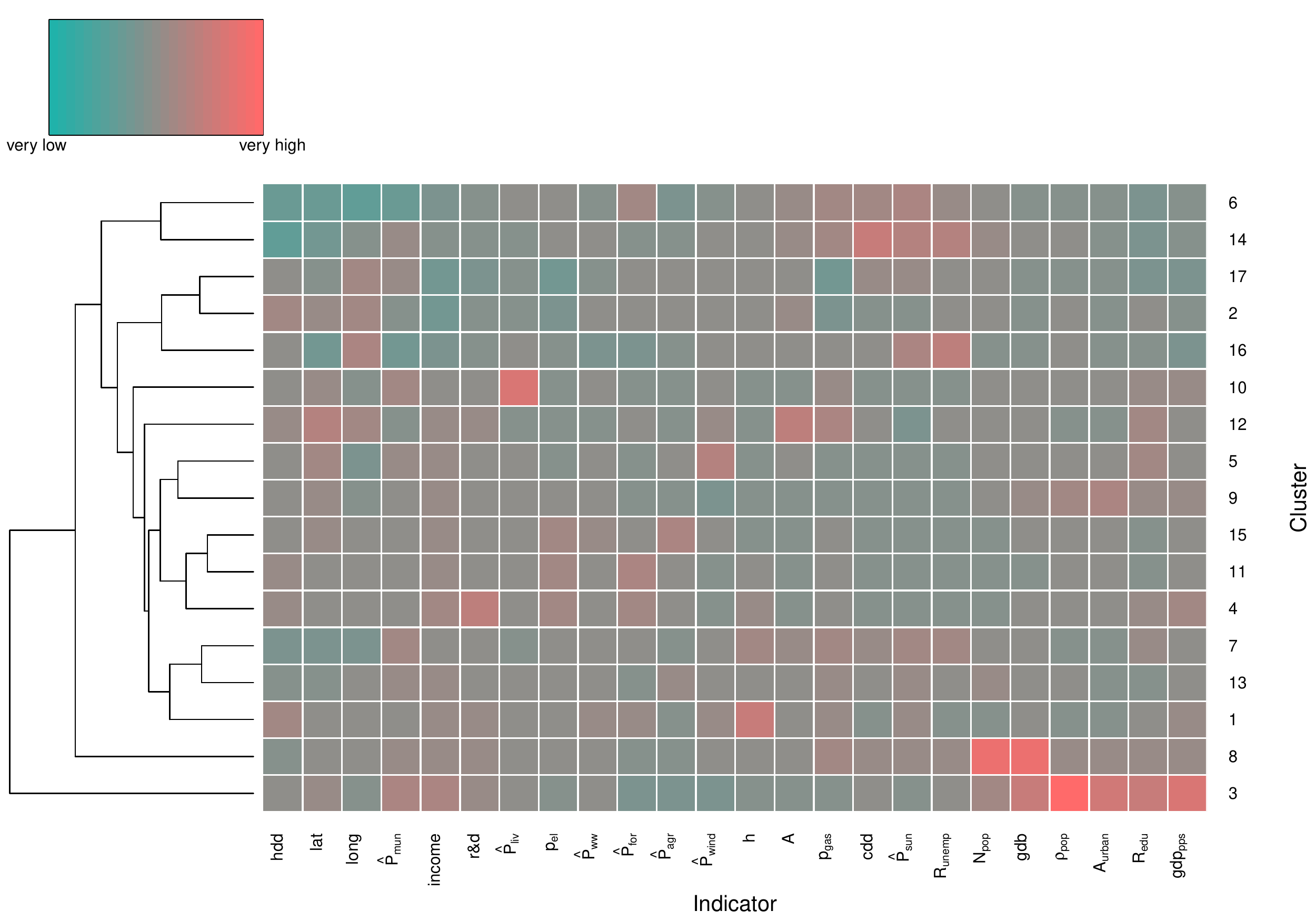}
\caption{Centroids analysis: the heat map shows high and low values for the variables in each centroid, while the dendrogram groups similar clusters based on the values assumed by the variables in the centroid.}
\label{figure_2}
\end{figure}

The \textit{k-means} algorithm grouped the territories of EU28 and Switzerland into 17 clusters.
Each cluster is characterized by a centroid, representing the mean value of each variable, for all elements of the cluster.
The centroids of all clusters were analysed to identify possible correlations. 
Figure \ref{figure_2} highlights the correlations among clusters through a heat map coupled with a dendrogram. 
The heat map shows the dimension of the variables for each cluster centroid through different colors: red shades indicate high values while aquamarine shades indicate low values. 
The dendrogram groups the clusters according to the presence of similar values for a significant number of variables.
For instance, clusters 2 and 17 belong to the same branch of the dendrogram since they show similar values for almost all variables.

\subsection{Classification of territories of EU28 and Switzerland}

\begin{figure}[!ht]
\centering\includegraphics[width=1\linewidth]{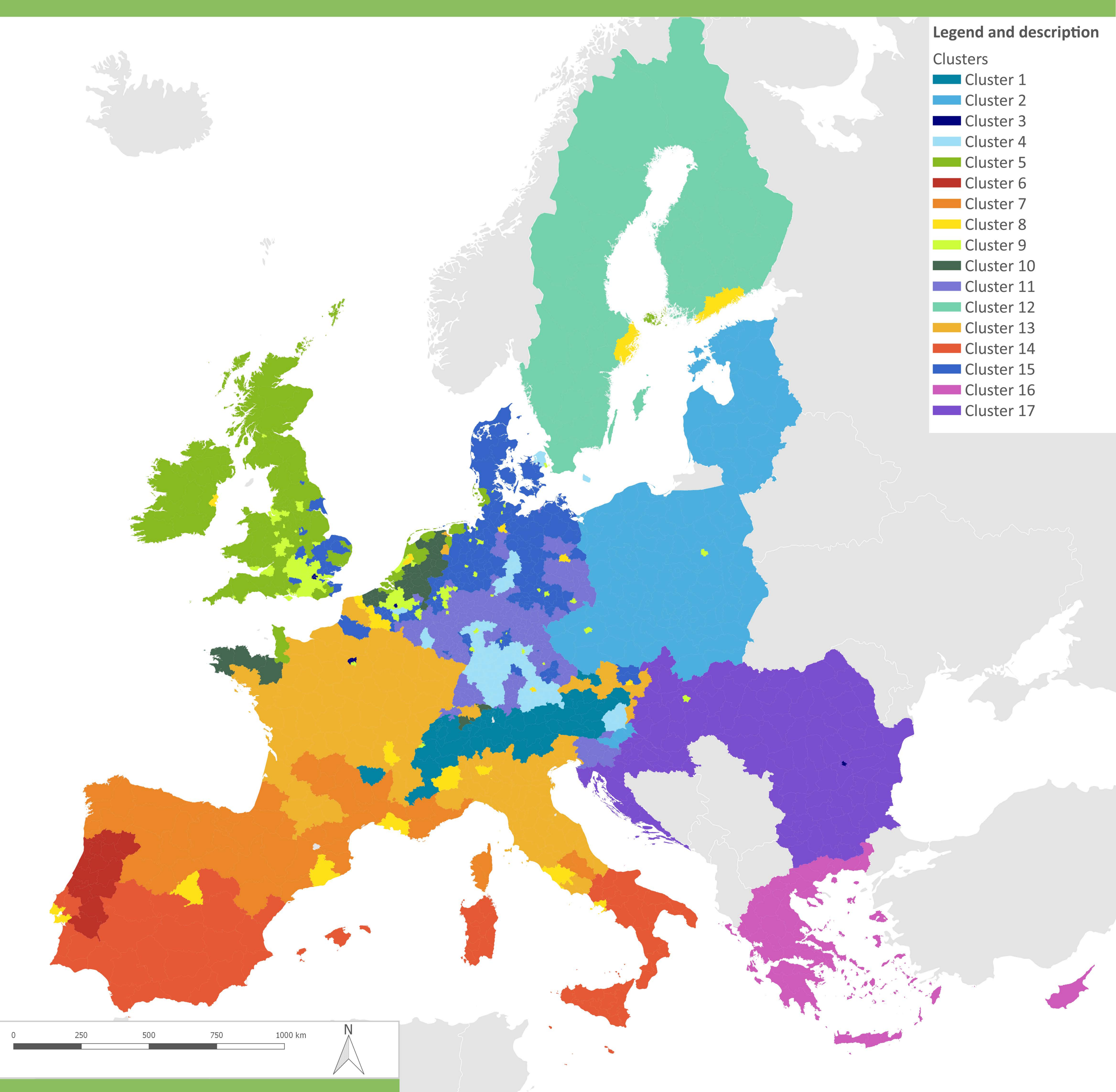}
\caption{Classification of territories of EU28 and Switzerland according to the k-means clustering algorithm. Territories showing the same color, belong to the same cluster.}
\label{figure_3}
\end{figure}

The classification of territories is better visualized through the map in Figure \ref{figure_3}.
After a first analysis of cluster centroids, the quantitative values assumed by the indicators in the centroids have been converted into qualitative assessments, so that in the centroid of each cluster, the indicator can assume one of the following values: \textit{very high, high, medium/high, medium, medium/low, low, very low}. 

Below are listed the most relevant clusters in terms of renewable energy potential, i.e. the ones presenting \textit{very high} or \textit{high} values for at least one indicator relative to renewable energy potential. 
For each cluster, the other renewable source or sources with a high potential indicator are mentioned, along with the values assumed by non-energy indicators, when significant. 
Examples of territories comprised in each cluster are given.

Clusters 2, 9, 12, 13 and 17 are not mentioned due to the low values assumed by energy indicators in their centroid.

\subparagraph{Cluster 1 - Alpine areas with very high potential from waste water per person} This cluster includes the Alpine area, encompassing the southern part of Austria, four departments of southeastern France, six provinces of northern Italy and the best part of Switzerland. 
From an energy point of view, the elements of this cluster present also a \textit{medium/high} potential from forest biomass, a \textit{medium/high} potential from municipal solid waste per person and a \textit{medium} potential from wind. 
Elevation in the territories in this area is \textit{very high}, consequently the heating demand assumes \textit{very high} values while the cooling demand assume \textit{very low} values. 
For what concerns socioeconomic indicators, the unemployment rate is \textit{very low} and so is GDP in PPS per person.

\subparagraph{Cluster 3 - Highly urbanized areas with very high potential from municipal solid waste per person} This cluster includes the metropolitan boroughs of the inner area of London (UK), the city of Paris (France) and two other departments of the Ile-de-France, the \textit{arrondissement} (district) of Bruxelles (BE) and the Romanian capital of Bucharest. 
From an energy point of view, the elements of this cluster also present a \textit{medium/high} potential from waste water per person. 
The heating demand is \textit{medium/high}.
For what concerns socioeconomic indicators, population density and concentration of urbanized areas are \textit{very high}, the unemployment rate is \textit{low}, income per capita and GDP in PPS per capita are both \textit{very high}.
As for education and research, these areas show a \textit{very high} percentage of people with tertiary education. 

\subparagraph{Cluster 4 - Innovation oriented areas with high potential from forest biomass per m\textsuperscript{2} and high potential from waste water treatment plants per person} This cluster includes mostly German territories, like the \textit{Landkreis} of Dachau; it also comprises the \textit{arrondissement} of Nivelles (Belgium), the region of North Zealand and the island of Bornholm (Denmark), the state of Luxembourg and three districts of the Bundesland of Styria (Austria). 
From an energy point of view, the elements of this cluster present also a \textit{medium/high} potential from municipal solid waste per person
Electricity prices are \textit{very high} and the heating demand is \textit{high}.
For what concerns the socioeconomic indicators, the unemployment rate is \textit{very low} and income per capita is \textit{high}.
This is the only cluster where the expenditure for research and development per capita assumes \textit{very high values}.

\subparagraph{Cluster 5 -  North Atlantic areas with very high potential from wind energy} This cluster includes mostly territories from the UK and Ireland, like the counties of Durham (UK) and Kerry (Ireland); it also comprises five territories in northern Germany, the \AA land Islands in Finland, the French department of Manche and nine territories in the northern Netherlands. 
From an energy point of view, the elements of this cluster present also a \textit{medium/high} potential from municipal solid waste per person.
The heating demand is \textit{high}.
For what concerns the socioeconomic indicators, GDP in PPS per person is \textit{low}, while the unemployment rate is \textit{very low}.

\subparagraph{Cluster 6 - South Atlantic areas with very high potential from forest biomass per m\textsuperscript{2} and high potential from solar energy} This cluster includes the best part of Portugal and the French overseas departments of Guyane, Guadeloupe and Martinique. 
From an energy point of view, the elements of this cluster present \textit{high} gas prices. 
For what concerns the socioeconomic indicators, GDP in PPS per person is \textit{very low} and income per capita is \textit{low}.
As for education and research, these areas show a \textit{very low} percentage of people with a tertiary education and a \textit{very low} expenditure in research and development per capita. 

\subparagraph{Cluster 7 - Mountain areas with high potential from municipal solid waste per person} This cluster includes the northern region of Spain including the territories of the Pyrenees, several southern mountainous regions of France including the island of Corsica, the Canary Islands (Spain), the intermunicipal communities of Cávado and Alto Minho and the archipelago of Madeira (Portugal), and two provinces of central Italy. 
From an energy point of view, the elements of this cluster present a \textit{medium/high} value for solar energy potential and suffer \textit{high} gas prices. 
For what concerns the socioeconomic indicators, GDP in PPS per capita is \textit{low}.

\subparagraph{Cluster 8 - Highly populated areas with high potential from municipal solid waste per capita} This cluster includes capital cities and highly populated urban areas of Europe: the city of Lisbon (Portugal), the cities of Madrid and Barcelona (Spain), the cities of Rome, Milan, Turin and Naples (Italy), the departments of Bouches-du-Rhône, with the city of Marseille, Rhône-Alpes, with the city of Lyon, and Nord, with the city of Lille (France), the city of Dublin (Ireland), the city of Amsterdam (Netherlands), the cities of Berlin, Hamburg and Munich (Germany), the city of Wien (Austria), the city of Stockholms (Sweden) and the city of Helsinki (Finland).
From an energy point of view, the elements of this cluster suffer a \textit{high} price for gas. 
For what concerns the socioeconomic indicators, population and GDP of these areas are \textit{very high}. 

\subparagraph{Cluster 10 - North Atlantic areas with a very high potential from livestock effluents per m\textsuperscript{2} and high potential for municipal solid waste per person} This cluster includes territories commonly known for livestock breeding: the best part of the Netherlands, the West Flanders territories in Belgium, two Swiss cantons and three departments in the French region of Brittany. 
For what concerns the socioeconomic indicators, the elements of this cluster present a \textit{very low} unemployment rate and a \textit{low} GDP in PPS per capita.

\subparagraph{Cluster 11 - Central European areas with very high potential from forest biomass per m\textsuperscript{2} and high potential from waste water per person} This cluster includes mostly German and Belgian territories, like the \textit{Landkreis} of Ostallgäu and the \textit{arrondissement} of Verviers; it also comprises two Swiss cantons and the best part of Slovenia. 
From an energy point of view, the elements of this cluster suffer \textit{very high} prices for electricity and present a \textit{high} heating demand.
For what concerns the socioeconomic indicators, GDP in PPS per person is \textit{low}, while the unemployment rate is \textit{very low}.
As for education and research, the percentage of people with a tertiary education is \textit{low} and so is the expenditure in research and development per capita.

\subparagraph{Cluster 14 - West Mediterranean areas with high potential from municipal solid waste per capita and very high potential from solar energy} This cluster comprises all territories of southern Italy, including the islands of Sicily and Sardinia, all southern Spain territories, two inter-municipal communities of Portugal and the state of Malta. 
From an energy point of view, the elements of this cluster present a \textit{medium/high} potential from waste water and a \textit{medium} potential from wind energy. 
They have a \textit{very high} cooling demand, and suffer \textit{very high} gas prices. 
For what concerns the socioeconomic indicators, GDP in PPS is \textit{very low} and the unemployment rate is \textit{high}.
As for education and research, a \textit{very low} percentage of people have tertiary education and the expenditure for research and development per capita is \textit{very low}.

\subparagraph{Cluster 15 - North Atlantic areas with very high potential from agricultural biomass per m\textsuperscript{2} and high potential from waste water per person} This cluster includes territories of north and central Europe commonly known for their agricultural production, encompassing numerous areas of Germany, the French department of Somme, part of the eastern cost of England (UK), the best part of Denmark, ten Belgian territories and two regions of northern Austria. 
The elements of this cluster suffer \textit{very high} prices for electricity and present a \textit{high} heating demand.
For what concerns the socioeconomic indicators, the unemployment rate is \textit{very low}, and GDP in PPS per capita is \textit{low}.
As for education and research, a \textit{low} percentage of people have tertiary education and the expenditure for research and development per capita is \textit{low}.

\subparagraph{Cluster 16 - East Mediterranean areas with high potential from solar energy}
This cluster includes all continental territories and islands of Greece and the state of Cyprus. 
For what concerns the socioeconomic indicators, the elements of this cluster present a \textit{very high} unemployment rate, a \textit{low} income per capita, a \textit{very low} GDP in PPS per capita.
As for education and research, a \textit{low} percentage of people have tertiary education and the expenditure for research and development per capita is \textit{very low}.

\subsection{Comparison between EU macro-regions and clusters: heterogeneities and common grounds}

As it can be seen in the following maps, the macro regions of EU, often object of energy and non-energy related projects, encompass heterogeneous territories from the point of view of the classification here performed. 
It is therefore essential to understand what are the causes for this heterogeneity and what the territories comprised in the macro-areas have in common beside geographic location, in order to encourage more effective policies.

\begin{figure}[!ht]
\centering\includegraphics[width=1\linewidth]{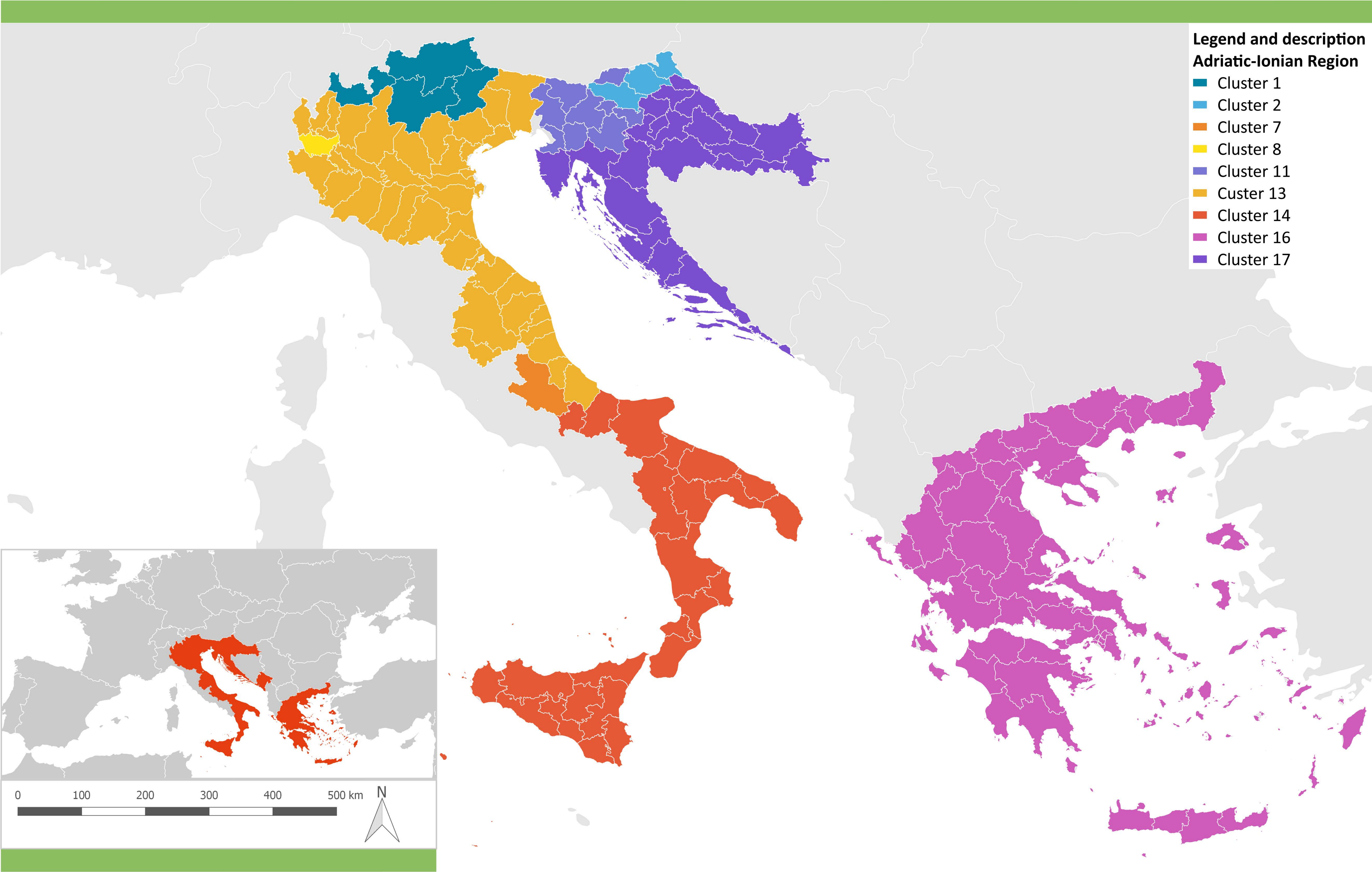}
\caption{Classification of territories of the Adriatic-Ionian macro region according to the cluster analysis}
\label{figure_4}
\end{figure}

The Adriatic-Ionian region (Figure \ref{figure_4}) includes territories belonging to nine clusters inside its boundaries. 
For what concerns the assessment of renewable energy sources, the territories of this area present a similar potential from municipal solid waste, generally assuming high values, while all other sources present quite diverse potentials.
The energy market is very heterogeneous: electricity and gas have very diverse prices according to the country. 
The potential energy demand, under the hypothesis of being correlated with heating and cooling degree days, assumes very different values, ranging from \textit{very low} to \textit{very high}. 
The economic context also presents some inconsistencies. 
Unemployment rate is generally low with the exception of cluster 14 and 16, presenting respectively \textit{high} and \textit{very high} values. 
Income per capita assumes very different values while the GDP in PPS per capita is generally low.
As for research and education, the territories share a low percentage of people with tertiary education as well as a low expenditure for research and development.   

\begin{figure}[!ht]
\centering\includegraphics[width=1\linewidth]{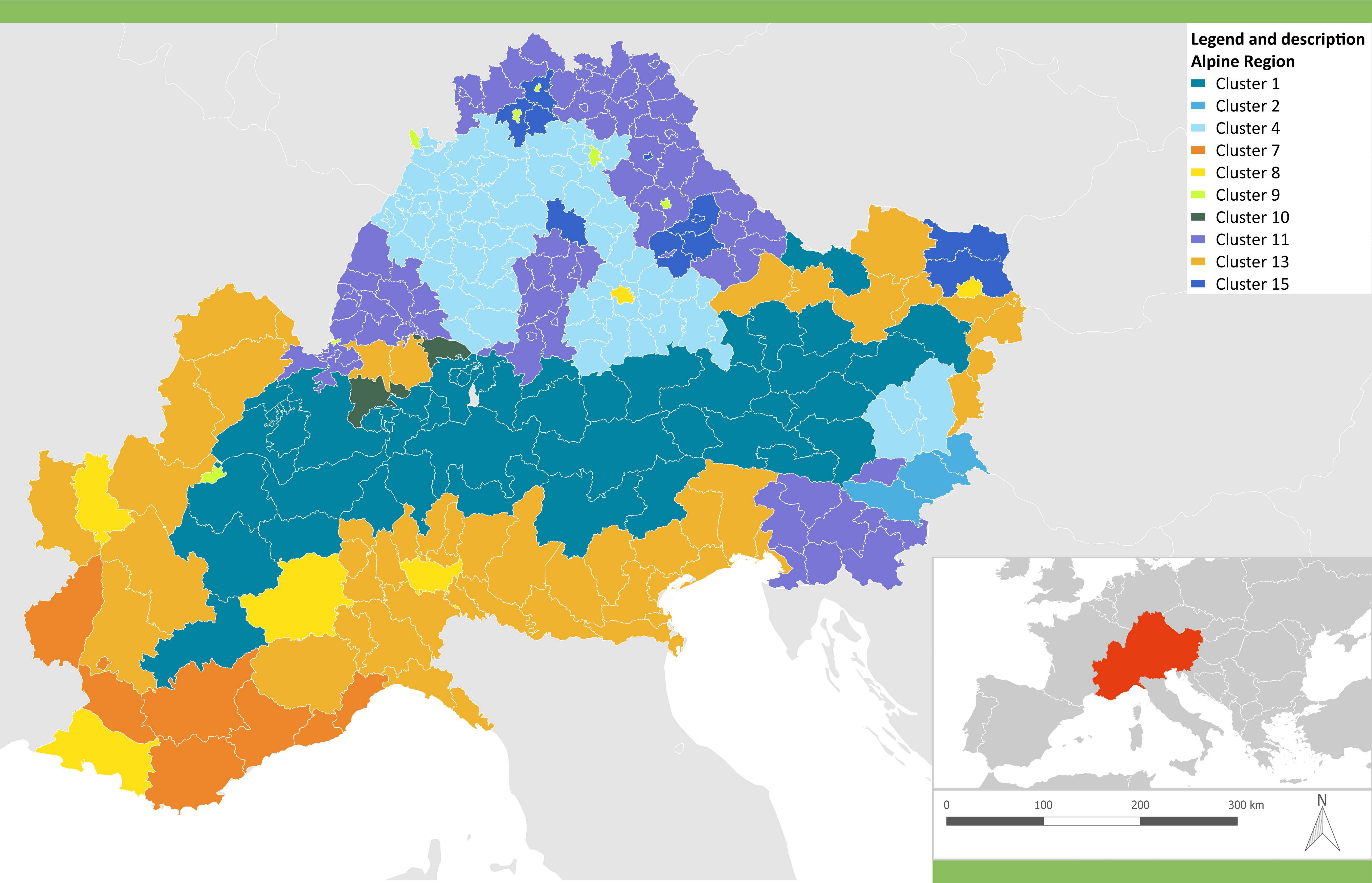}
\caption{Classification of territories of the Alpine macro region according to the cluster analysis}
\label{figure_5}
\end{figure}

The Alpine region (Figure \ref{figure_5}) includes territories belonging to ten clusters inside its boundaries. 
For what concerns the assessment of renewable energy sources, the territories of this area present a similar potential from municipal solid waste and from waste water, assuming mainly medium and high values; all other sources though present very diverse potentials. 
The prices of electricity and gas also assume very different values.
Conversely, the heat demand is homogeneous in the area due to the mountain environment covering a considerable share of the territories: the values for heating degree days is generally high, while that for cooling degree day is low.
The economic context is quite consistent among the territories: unemployment is generally low, income per capita assumes mainly high values while the GDP in PPS per capita is low. 
As for research and education, the territories share a low-medium/low percentage of people with tertiary education but the expenditure in research and development range from \textit{very low} to \textit{very high} values.  

\begin{figure}[!ht]
\centering\includegraphics[width=1\linewidth]{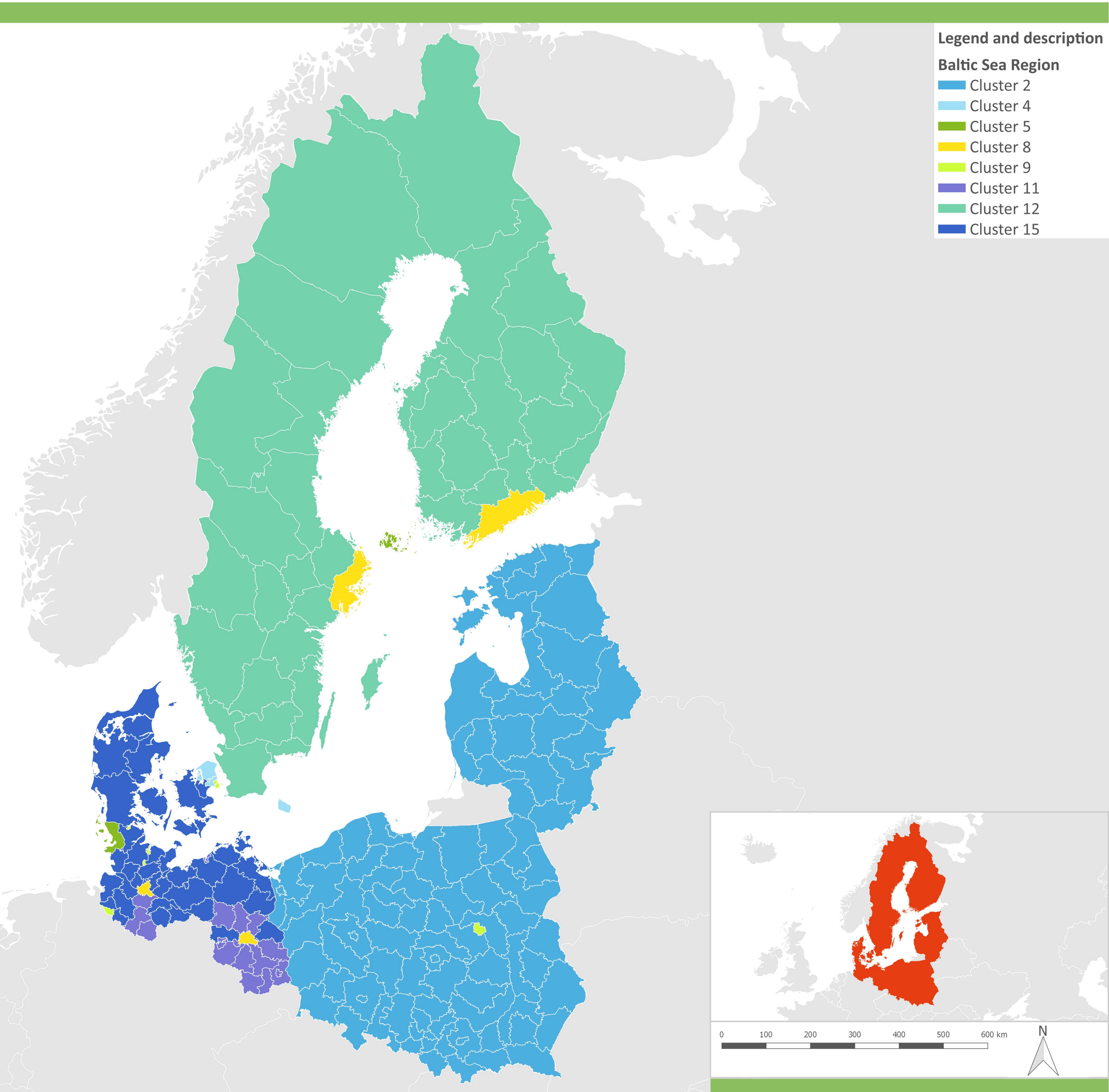}
\caption{Classification of territories of the Baltic Sea macro region according to the cluster analysis}
\label{figure_6}
\end{figure}

The Baltic Sea region (\ref{figure_6}) includes territories belonging to eight clusters inside its boundaries. 
For what concerns the potential from renewable energy sources, the territories of this area are quite similar, in particular the potential from municipal solid waste and from waste water is medium/high while the majority of the others are low. 
There are though two exceptions: the potentials from forest biomass and wind power both range from \textit{very low} to \textit{very high} values. 
The prices of electricity and gas assume very different values. 
Conversely, the heat demand is very similar in the whole area: heating degree days assume high values while cooling degree days are generally low, due to the high latitude of the territories. 
The economic context is quite consistent among the territories: unemployment is low, income per capita mainly assumes medium/high values and the GDP in PPS per capita is low. 
As for research and education, the territories share a low percentage of people with tertiary education while expenditure in research and development assumes very different values.

\begin{figure}[!ht]
\centering\includegraphics[width=1\linewidth]{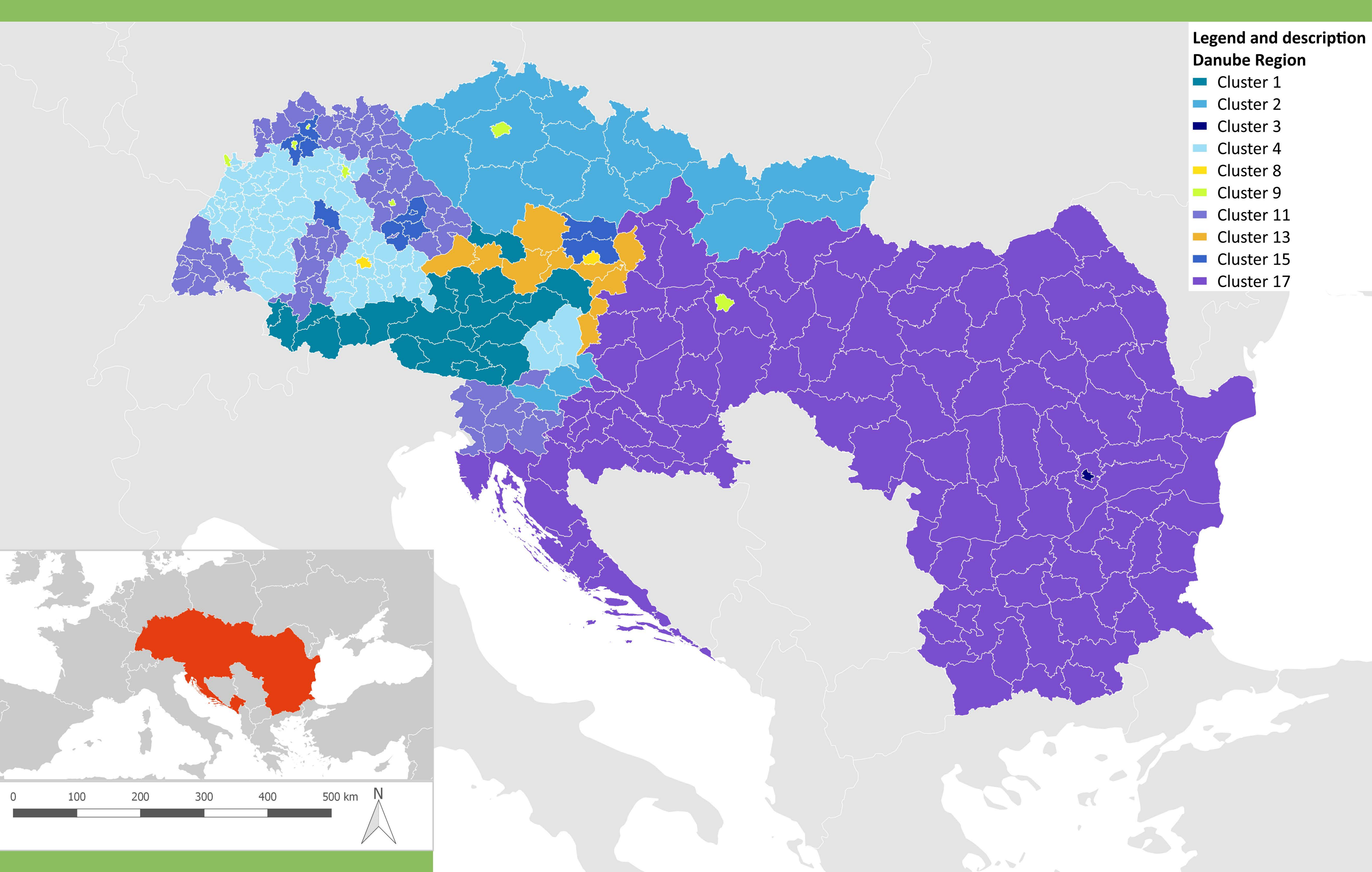}
\caption{Classification of territories of the Danube macro region according to the cluster analysis}
\label{figure_7}
\end{figure}

The Danube region (\ref{figure_7}) includes territories belonging to ten clusters inside its boundaries. 
For what concerns the potential from renewable energy sources, the territories of this area are quite different with the exception of the potentials from municipal solid waste and waste water, generally high.   
Conversely, the heat demand is consistent in the whole area: heating degree days are usually high while cooling degree days are low. 
The energy context present heterogeneity regarding the prices of electricity and gas, both ranging from \textit{very low} to \textit{very high} values. 
The economic context is heterogeneous when it comes to GDP in PPS per capita, instead unemployment and income are consistent, assuming respectively low and medium/high values. 
The region presents pronounced differences in the research and education context: both the percentage of people with tertiary education and the expenditure in research and development range from \textit{very low} to \textit{very high} values.\\

Even though dissimilarities are present in European macro-region, the clusters obtained in this study highlight similarities among territories in different member states that could strengthen territorial cohesion and constitute a base for further EU energy project and policies. 
Territories with similar energy potential can lay the groundwork for cooperation, exchange good practices and policies and encourage renewable energy production.

\begin{figure}[!ht]
\centering\includegraphics[width=1\linewidth]{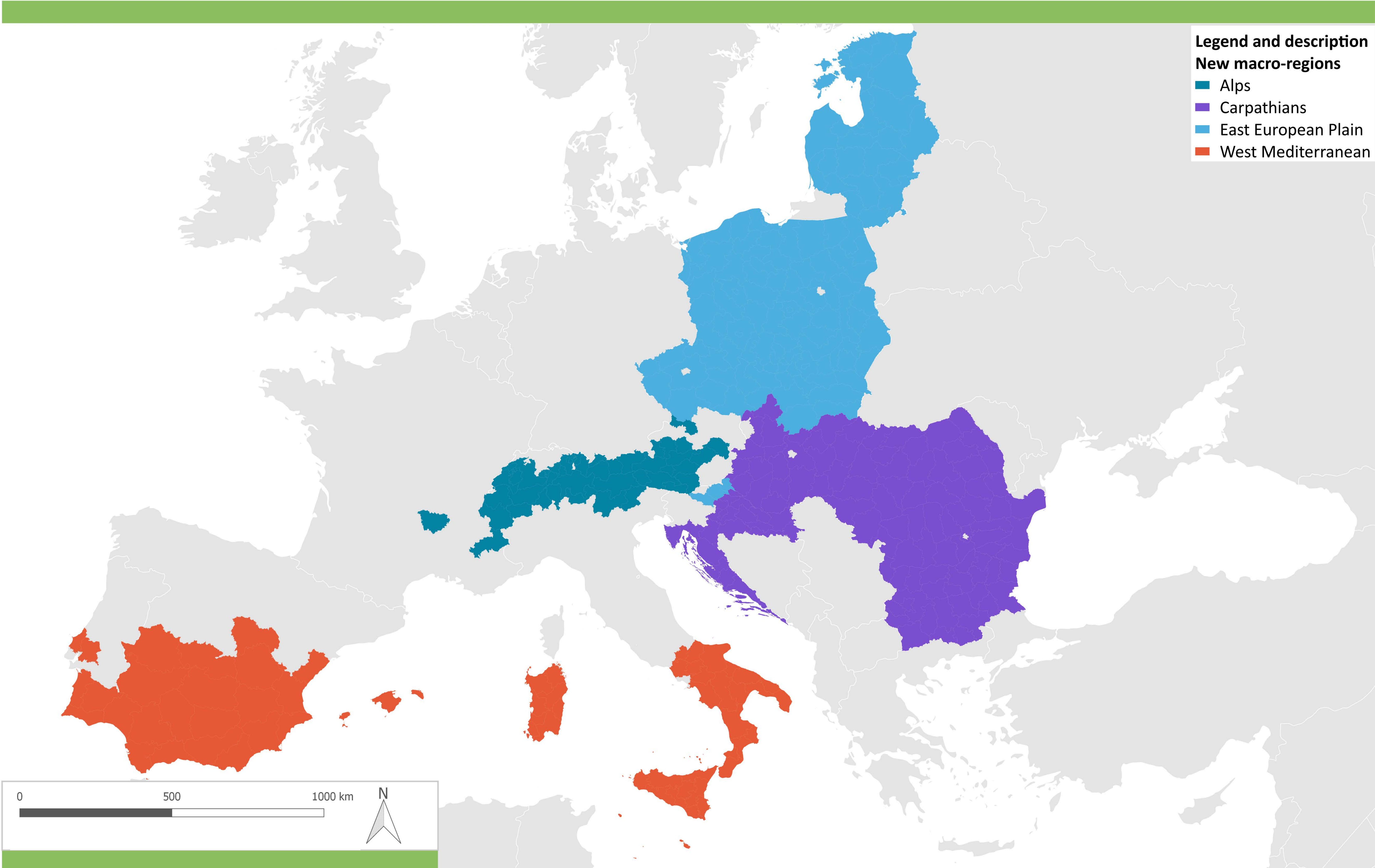}
\caption{Possible new macro-regions based on the boundaries of clusters 1 (Alpine territories), cluster 14 (West Mediterranean territories), cluster 2 (East European territories) and cluster 17 (Carpathian territories).}
\label{figure_8}
\end{figure}

Four examples of clusters with the potential of becoming macro-regions are represented in Figure \ref{figure_8} and listed as follows:
\begin{itemize}
\item The Alps, stretching from Switzerland to Austria, including territories of France and northern Italy (cluster 1)
\item The West Mediterranean, encompassing southern Italy and its islands, and the south of Spain with the Balearic islands (cluster 14)
\item The East European Plain, including Poland, Czech Republic and the Baltic countries (cluster 2) 
\item The Carpathians, comprising Bulgaria, Romania, Hungary, Slovakia and Croatia (cluster 17)
\end{itemize} 

\section{Conclusions and policy recommendations}
\label{conclusion}

The analysis here performed lays the foundations for a comprehensive territorial classification of the EU28 area and Switzerland, one that includes details on the energy profile, the geomorphological characterization and the socio-economic structure of the territories. 
Two key aspects are stressed by means of this classification:
\begin{itemize}
\item the study area is very heterogeneous: territories located in the same country or geographical region can present substantial differences in energy and non-energy related aspects; they thus face different challenges and issues when it comes to renewable energy production and consequently require different policy approaches
\item although heterogeneous, the study area presents interesting spatial patterns: territories sharing neither national identity nor similar geographic coordinates can have a lot in common; from an energy perspective they can hold comparable sources or be subject to the same energy prices, as for the non-energy layout, they could have developed similar economic environments or be interested by analogous social phenomena. 
\end{itemize}
The inclusion in the analysis of both energy and non-energy related indicators provides a new perspective on how to approach cross-country policies and programmes.
Specifically, two types of policy recommendation arise from these results.

At macro level, this classification can be used by EU decision makers for designing more focused energy policies. 
The scope of macro-regional policies could be adjusted according to the clusters resulting from this study, in order to obtain tailor-made and territory oriented measures.
For instance, clusters 1, 7 and 10 are the core of the Alpine macro-region (Figure \ref{figure_5}); these clusters are in fact statistically "close" when looking at the dendrogram distribution in Figure \ref{figure_2}, meaning they behave similarly with respect to relevant indicators. From an energy perspective, the potential from municipal solid waste, electricity and gas prices assume similar values; unemployment in generally low, income per capita is medium and GDP in PPS per capita is low in all three clusters. Nevertheless, other indicators assume very different values and projects should focus on leveling out these differences in order to use territorial cohesion as a mean for promoting renewable energy production.

At micro level, local administrators and energy managers can use these results as an energy planning tool, by comparing their experience with that of other territories belonging to the same cluster and formulating energy related plans accordingly. 
The territories of a cluster could exchange ideas and good practices among each other, try out successful strategies, and avoid unsuccessful ones, \textit{ipso facto} facilitating decision making and turning energy planning into a bottom-up process.
For instance, all cities comprised in cluster 8 share a high potential from municipal solid waste, though some managed to put in place more successful strategies for producing energy from this renewable source, than others. 
The exchange of knowledge and information could be key in spreading effective measures in this sense, in all territories of the cluster.

Nevertheless, limits to this method and study exist. 
At macro level, the analysis carried out in this study is static, i.e. the data on which the cluster analysis is based on, are not updated. 
This issue was partially solved by accurately explaining the method: this ensure replicability. 
The classification of territories can be repeated based on updated data and give new results and insights.
Still, it requires a huge statistical and research effort to recalculate and reassess all energy and non-energy indicators at NUTS3 level for all EU28 and Switzerland in order to provide more valid and up-to-date data.

At micro level, the biggest limit is that this tool cannot be used by policy makers at local level to design the actual implementation of renewable energy production plants. 
This classification is intended to serve as a support tool to draft strategies and start collaborations among close and distant territories, in order to use energy as a catalyst for territorial cohesion and make renewable energy planning more tailored and effective.

\section{Acknowledgment}
This study was conducted in the frame of the European Union (EU) H2020 project HotMaps (grant number: \href{https://cordis.europa.eu/project/rcn/205761/factsheet/en}{723677}) project. 
The Horizon 2020 \href{https://www.hotmaps-project.eu/}{HotMaps} project aims at developing "an open source heating/cooling mapping and planning toolbox" for EU28 policy makers at national and local level \cite{noauthor_hotmaps_nodate}. 
The authors would like to thank all the partners involved in the project and especially the \href{https://eeg.tuwien.ac.at/staff/people/lukas-kranzl}{Technical University of Wien (Institute of Energy Economics)} for the coordination and support in all the task activities.

\section{References}
\label{reference}

\bibliography{biblio}

\end{document}